\documentclass[a4paper,12pt]{article}
\usepackage[english]{babel}
\usepackage{graphicx,color}
\usepackage{amsmath,amsthm,amssymb,amsfonts}
\usepackage{epsfig}
\usepackage[T1]{fontenc}
\usepackage[cp1250]{inputenc}
\usepackage[bookmarks,colorlinks]{hyperref}
\usepackage{fancyhdr}
\fancypagestyle{plain}{%
\fancyhead{}

}

\hypersetup{colorlinks,%
linkcolor=magenta,%
urlcolor=blue,%
citecolor=red}

\title{New Approach to Quantization of Cosmological Models}

\author{\textbf{\L ukasz Andrzej Glinka}\vspace*{15pt}\\
E-mail: \href{mailto:laglinka@gmail.com}{\bf{\tt{laglinka@gmail.com}}}
\vspace*{10pt}\\
\emph{International Institute for Applicable}\\
\emph{Mathematics \& Information Sciences,}\\
\emph{Hyderabad (India) \& Udine (Italy),}\vspace*{10pt}\\
\emph{B.M. Birla Science Centre,}\\
\emph{Adarsh Nagar, 500 063 Hyderabad, India}
}
\date{\today}

\begin{document}
\maketitle
\thispagestyle{empty}
\begin{abstract}
We propose the new quantization of homogenous cosmological models.
Four fundamental methods are applied to the cosmological model and efficiently jointed.
The Dirac method for constrained systems is used, then the Fock
space is built and the second quantization is carried out. Finally,
the diagonalization ansatz, which is a combination of the Bogoliubov
transformation method and the Heisenberg equation of motion, is formulated.
The temperature of quantum cosmological model is introduced.\\

\end{abstract}
\newpage
\section{\label{sec:1}Introduction}

As very well known, the quantum formulation of the general
relativity theory yields many problems. Admittedly exist a lot of
attempts of developing the quantum theory of gravitational field
and, hence, the quantum cosmology, but still we have no a good, in
mathematical and physical senses, a quantum theory (see e.g.
\cite{E} for review of quantum cosmology)

The most famous approaches to quantum cosmology, as \emph{e.g.}
canonical quantization, loop quantization, or path integral
quantization, in fact are based on too straightforward applications
of quantum mechanics. As a result, all above approaches yield the mathematical problems
and the physical aspects still are not clear. For example, in the
Wheeler--DeWitt equation based on the Hamiltonian constraint
following from $3+1$ space-time splitting, the functional
differential evolution is presented, what does not allow to solve
the generic problem in principle. Similar problems take place
manifestly in the functional integrals formulations, where the
conception of the wave function of the Universe did not get a clear
physical interpretation. From the other side loop quantum cosmology
results in self-inconsistency with experimental data.

In this paper we offer a new approach to the problem of formulating
the quantum cosmology. We begin with the homogeneous cosmological
model and develop a quantization procedures allowing to avoid some
of the existing problems by leaving quantum mechanics approach for
quantum field theory one, at least for the simple model. We propose
an effective combination of four elementary rules of quantization and as a
result we obtain well-defined, in quantum-field-theoretical sense,
the new model of quantum cosmology.

The paper is organized as following. In Section 2 we consider the
basic elements of the homogenous cosmological model, which will be
used in the next Sections. Section 3 is devoted to presentation of
the quantization method, that mixes non trivially the Dirac primary
quantization of constrained systems and the Fock second quantization
with the Bogoliubov diagonalization method. In the Section 4 we
discuss the possibly simplest model for thermodynamic interpretation
of quantum cosmology, that leads to the concept of temperature
associated with the quantum cosmological model. In the Section 4 a
brief summary of the results is done.

\section{\label{sec:2} Cosmological Model}

Dynamics of any space-time in general relativity is defined by the action depending on
metric tensor and matter fields
\begin{equation}\label{2.1}
\mathcal{A}=\int d^{4}x \sqrt{-g}
\left(-\frac{1}{2\kappa}\mathcal{R}+\mathcal{L}_{Fields}\right),
\end{equation}
where $g=\det{g_{\mu\nu}}$, $g_{\mu\nu}$ is the metric tensor,
$\kappa=8\pi G$, $G$ is the Newton gravitational constant,
$\mathcal{L}_{Fields}$ is a Lagrangian density of the matter fields,
\mbox{$\mathcal{R}=g^{\mu\nu}R_{\mu\nu}$} is the Ricci curvature
scalar and
\begin{eqnarray}
R_{\mu\nu}&=&\partial_{\alpha}\Gamma^{\alpha}_{\mu\nu}-\partial_{\nu}\Gamma^{\alpha}_{\mu\alpha}+\Gamma^{\alpha}_{\beta\alpha}\Gamma^{\beta}_{\mu\nu}-\Gamma^{\alpha}_{\beta\nu}\Gamma^{\beta}_{\mu\alpha},\label{2.3}\\
\Gamma^{\rho}_{\mu\nu}&=&\frac{1}{2}g^{\rho\sigma}\left(\partial_{\nu}g_{\mu\sigma}+\partial_{\mu}g_{\sigma\nu}-\partial_{\sigma}g_{\mu\nu}\right),\label{2.4}
\end{eqnarray}
are the Ricci curvature tensor and the Christoffel affine connections respectively.
The gravitational equations of motion following from this action have the form
\begin{equation}\label{2.5}
R_{\mu\nu}-\frac{1}{2}\mathcal{R}g_{\mu\nu}=\kappa T_{\mu\nu},
\end{equation}
where
\begin{equation}\label{2.6}
T_{\mu\nu}= \frac{2}{\sqrt{-g}} \frac{\delta
\mathcal{A}_{Fields}}{\delta g^{\mu\nu}}\quad,\quad\mathcal{A}_{Fields}\equiv\int d^4x\sqrt{-g}\mathcal{L}_{Fields}
\end{equation}
is the energy-momentum tensor of the matter fields \cite{W,Mis}. Here
$\mathcal{A}_{Fields}$ is the matter fields action.

Let us consider the homogenous and isotropic cosmological model with
metric \cite{Mis}:
\begin{equation}\label{3.1}
g_{\mu\nu}=\left[\begin{array}{cccc}N_{d}^{2}(t)&0\\0&-a^{2}(t)\delta_{ij}\end{array}\right],
\end{equation}
where $N_{d}$ is the Dirac lapse function, and $a$ is the Friedmann
cosmological scale factor. Non-vanishing Christoffel symbols
(\ref{2.4}) for the metric tensor (\ref{3.1}) are:
\begin{equation}\label{3.5a}
\Gamma^{0}_{00}=\frac{\dot{N}_{d}}{N_{d}},~~\Gamma^{0}_{ii}=\frac{a\dot{a}}{N^{2}_{d}},~~\Gamma^{i}_{i0}=\frac{\dot{a}}{a},
\end{equation}
where $i=1,2,3$ is space-like index. By direct using of (\ref{3.5a}) one obtains the Ricci tensor
(\ref{2.3}) in the form:
\begin{equation}\label{3.6}
R_{\mu\nu}=\left[\begin{array}{cccc}-3\left(\frac{\ddot{a}}{a}-\frac{\dot{a}\dot{N_{d}}}{aN_{d}}\right)&\mathbf{0}^{\mathrm{T}}\\
\mathbf{0}&\left(\frac{a\ddot{a}}{N^{2}_{d}}-2\frac{\dot{a}^{2}}{N^{2}_{d}}-\frac{a\dot{a}\dot{N}_{d}}{N^{3}_{d}}\right)\delta_{ij}\end{array}\right],
\end{equation}
where  $\mathbf{0}$ is a null vector. In this manner one establishes the curvature scalar as
\begin{equation}
\mathcal{R}=-6\left(\frac{\ddot{a}}{aN^{2}_{d}}-\frac{\dot{a}^{2}}{a^{2}N^{2}_{d}}-\frac{\dot{a}\dot{N}_{d}}{aN_{d}^{3}}\right).
\end{equation}

We introduce the conformal time $\tau$ and the scale of masses
$\phi$ by the formulae
\begin{eqnarray}
\tau(t)&=&\int_{0}^{t}\frac{N_{d}(t')}{a(t')}dt',\label{3.2}\\
\phi(\tau)&=&\sqrt{\frac{3}{8\pi}}\mathcal{M}_{Pl}a(\tau),\label{3.3}
\end{eqnarray}
where $\mathcal{M}_{Pl}=\frac{1}{\sqrt{G}}$ is the Planck mass in
the natural units. In the conformal time (\ref{3.2}) the space-time
metric looks like pseudo-euclidian interval
\begin{equation}\label{3.4}
ds^{2}=a^{2}(\tau)\left\{d\tau^{2}-\left(dx^{2}+dy^{2}+dz^{2}\right)\right\}.
\end{equation}
Reduced by partial integration, the action (\ref{2.1}) expressed in
the terms of conformal time (\ref{3.2}) and the scale of masses
(\ref{3.3}) becomes
\begin{equation}\label{3.5}
\mathcal{A}[\phi]=\int
d\tau\left\{-V\frac{d\phi}{d\tau}\frac{d\phi}{d\tau}-\left(\frac{8\pi}{3\mathcal{M}_{Pl}^{2}}\right)^{2}\phi^{4}\mathrm{H}_{Fields}(\tau)\right\},
\end{equation}
where $V=\int_{\mathbb{R}^{3}}d^{3}x$ is spatial volume and
\begin{equation}
\mathrm{H}_{Fields}(\tau)=\int_{\mathbb{R}^{3}}d^{3}x\mathcal{H}_{Fields}(x,\tau)
\end{equation}
is the Hamiltonian of the Matter fields, where energy density $\mathcal{H}_{Fields}$ is received by using of the Legendre transformation to Lagrangian of matter fields $\mathcal{L}_{Fields}$. One derives the canonical conjugate momentum with respect to the action (\ref{3.5}) as
\begin{equation}\label{3.6}
P_{\phi}=\frac{\delta\mathcal{A}[\phi]}{\delta\left(d\phi/d\tau\right)}=-2V\frac{d\phi}{d\tau},
\end{equation}
and by application of the Hamiltonian reduction to (\ref{3.5}) one obtains
\begin{equation}\label{3.7}
\mathcal{A}[\phi]=-\int
d\tau\left\{P_{\phi}\frac{d\phi}{d\tau}-\mathrm{H}(\tau)\right\},
\end{equation}
where
\begin{equation}\label{3.8}
\mathrm{H}(\tau)=\frac{1}{4V^{2}}\left\{P_{\phi}^{2}-\left(\frac{16\pi
V}{3\mathcal{M}_{Pl}^{2}}\right)^{2}\phi^{4}\mathrm{H}_{Fields}(\tau)\right\}\equiv0
\end{equation}
is the shell-energy-constraint of the cosmological model. The Hamiltonian constraint
(\ref{3.8}) defines values of the canonical momentum (\ref{3.6}):
\begin{equation}\label{3.9}
P_{\phi}=\pm\mathrm{E}_{\phi},
\end{equation}
where we have introduced the quantity $\mathrm{E}_{\phi}$ defined by the formula:
\begin{equation}\label{3.10}
\mathrm{E}_{\phi}=\frac{16\pi
V}{3\mathcal{M}_{Pl}^{2}}\phi^{2}\sqrt{\mathrm{H}_{Fields}(\tau)}.
\end{equation}
The equations (\ref{3.6}) and (\ref{3.10}) allow determine one-dimensional the following evolution equation for the scale of mass $\phi(\tau)$
\begin{equation}\label{3.11}
\frac{d\phi(\tau)}{d\tau}=\pm\frac{8\pi}{3\mathcal{M}_{Pl}^{2}}\sqrt{\mathrm{H}_{Fields}(\tau)}\phi^{2}(\tau),
\end{equation}
which can be solve by straightforward using of elementary integration
\begin{equation}\label{3.12}
\pm\frac{8\pi}{3\mathcal{M}_{Pl}^{2}}\int_{\tau_{0}}^{\tau}\sqrt{\mathrm{H}_{Fields}(\tau')}d\tau'=\frac{1}{\phi(\tau)}-\frac{1}{\phi(\tau_{0})},
\end{equation}
with $\tau_0=\tau(t_0)$ as the initial value of the conformal time $\tau$. One can introduce into our considerations the redshift $z$
\begin{equation}\label{3.13}
z(\tau_{0};\tau)=\pm\frac{8\pi\phi(\tau_{0})}{3\mathcal{M}_{Pl}^{2}}\int_{\tau_{0}}^{\tau}\sqrt{\mathrm{H}_{Fields}(\tau')}d\tau',
\end{equation}
with using which the received solution (\ref{3.12}) takes more conventional form
\begin{equation}\label{3.14}
\frac{\phi(\tau)}{\phi(\tau_{0})}=\frac{1}{1+z(\tau_{0};\tau)},
\end{equation}
well known in literature as the Hubble law \cite{W}.

\section{\label{sec:3}Quantization}

The previous part of this paper was concentred on the some
properties of the cosmological model, which will be base for
discussion of its quantum aspects. We have considered a description of
the classical cosmology in terms of two parameters: the conformal time
(\ref{3.2}) and the scale of mass (\ref{3.3}). In
this section we are going to develop a quantum theory corresponding to this model.
To be more precise, we construct some "nonstandard" quantization
procedure being a nontrivial combination of famous elementary methods.
First we use \emph{the Dirac quantization} of constrained Hamiltonian systems, the next step is building the
\emph{the appropriate Fock space} and carrying out a secondary quantization in special form. Finally we formulate \emph{the Diagonalization Ansatz} as the combination of the Bogoliubov transformation and the Heisenberg equations of motion.

\subsection{\label{sec:4}First Quantization in the Dirac Approach}

The first step for building a quantum theory, is the first quantization of
energy constraints proposed by Dirac \cite{Dir} (Cf. also Ref. \cite{GT} where the method is worked out). The on-shell energy constraints of the classical theory
which we have deduced in the previous section are
\begin{equation}\label{4.1}
P_{\phi}^{2}-\mathrm{E}_{\phi}^{2}=0,
\end{equation}
where $\mathrm{E}_{\phi}$ is given by (\ref{3.10}) and $P_{\phi}$ is a canonical conjugate momentum of the classical cosmological model.
According to the Dirac approach, a first quantization of a classical theory is nothing else than quantization of the constraints (\ref{4.1})
by canonical commutation relation (CCR)
\begin{equation}\label{4.2}
i\left[P_{\phi},\phi\right]=1,
\end{equation}
and pure philosophical assuming about existence of a wave function of a quantum theory
$\Psi(\phi)$. This stage give us the wave equation associated to the quantum
cosmology:
\begin{equation}\label{4.3}
\left(\frac{\partial^{2}}{\partial\phi^{2}}+\mathrm{E}_{\phi}^{2}\right)\Psi(\phi)=0,
\end{equation}
which is here equivalent of well known the Wheeler--DeWitt equation \cite{WDW}, \cite{witt}. Note that the one-dimensional ($\phi$ is the dimension) evolution (\ref{4.3}) looks like formally as the  Klein-Gordon equation of massive scalar field, with time dependent mass, in $0$-dimensional space. This quantum cosmology has a global character manifestly.

\subsection{\label{sec:5}Fock Space and Second Quantization}

The next step of the quantization program is construction of the Fock space with correctly defined vacuum state and the second quantization for the quantum theory (\ref{4.3}). Naturally, by this step one obtains the description of the quantum cosmology in fundamental formalism of quantum field theory (Cf. \emph{e.g.} Refs. \cite{bog1, pesk, bog2}).

Let us consider the action functional $\mathcal{A}[\Psi]$ which should be use to obtaining of the equation (\ref{4.3}) as the Euler--Lagrange
equation of motion, according to the action principle $\delta \mathcal{A}[\Psi]=0$. By using of the naive analogy with the Klein--Gordon equation for a scalar massive particle, one observes that the action must have the following simple form
\begin{equation}\label{4.4}
\mathcal{A}[\Psi]=\int d\phi\left\{\frac{1}{2}\partial_{\phi}\Psi\partial_{\phi}\Psi-\frac{1}{2}\mathrm{E}_{\phi}^{2}\Psi^{2}\right\},
\end{equation}
for which the canonical conjugate momentum can be derived straightforwardly as
\begin{equation}\label{4.5}
\Pi_{\Psi}=\frac{\delta\mathcal{A}[\Psi]}{\delta\left(\partial_{\phi}\Psi\right)}=\partial_{\phi}\Psi.
\end{equation}
After the Hamiltonian reduction the proposed action (\ref{4.4}) takes the form
\begin{equation}\label{4.6}
\mathcal{A}[\Psi]=\int
d\phi\left\{\Pi_{\Psi}\partial_{\phi}\Psi-\frac{\Pi^{2}_{\Psi}+\mathrm{E}_{\phi}^{2}\Psi^{2}}{2}\right\}.
\end{equation}
In this manner we see that the Hamiltonian corresponding to the
action is established analogically to a simple harmonic oscillator's one
\begin{equation}\label{4.6a}
\mathrm{H}(\Pi_{\Psi},\Psi)=\frac{1}{2}\left(\Pi^{2}_{\Psi}+\mathrm{E}_{\phi}^{2}\Psi^{2}\right).
\end{equation}

We have seen that the wave equation of the quantum theory is in fact the Klein-Gordon evolution with changeable mass, which is locally (for fixed time) constant. By
this there is natural way for concluding that the Fock space of system should be boson type.
Let us build the Fock space of the annihilation and creation operators $\mathfrak{U}$ and
$\mathfrak{U}^{\dagger}$, which by the cosmological character of the quantum theory we will call name the Universe
annihilator and the Universe creator, respectively. The bosonic type CCRs are (Cf. \emph{e.g.} Ref. \cite{bog2})
\begin{equation}\label{4.7}
\left[\mathfrak{U},\mathfrak{U}^{\dagger}\right]=1,~~\left[\mathfrak{U},\mathfrak{U}\right]=0,~~\left[\mathfrak{U}^{\dagger},\mathfrak{U}^{\dagger}\right]=0.
\end{equation}
In this basis the canonical variables - the field $\Psi$ and its conjugate momentum $\Pi_{\Psi}$ - have the following Fock-like decomposition
\begin{eqnarray}
\Psi&=&\frac{1}{\sqrt{\mathrm{E}_{\phi}}}\frac{\mathfrak{U}^{\dagger}+\mathfrak{U}}{\sqrt{2}}\label{4.8},\\
\Pi_{\Psi}&=&i\sqrt{\mathrm{E}_{\phi}}\frac{\mathfrak{U}^{\dagger}-\mathfrak{U}}{\sqrt{2}}\label{4.9},\\
&&i\left[\Pi_{\Psi},\Psi\right]=1.
\end{eqnarray}
However, these distributions are not standard Fockian ones - the normalization
coefficients depend on the evolutionary dimension $\phi$. It must be emphasized that we introduced these coefficients~
in naive analogy with a harmonic oscillator case, and that the Fock-like bosonic operators $\mathfrak{U}$ and $\mathfrak{U}^{\dagger}$ create the Hilbert space. Because, this space is very special as well as is unknown evidently in literature, we propose to call name this space as the appropriate or appropriate Fock space. The similar situation is present in usual particle physics studies \cite{B}. In the appropriate Fock space of the Universe annihilators of creators (\ref{4.7}) the reduced theory action (\ref{4.6}) takes the form:
\begin{equation}\label{4.10}
\mathcal{A}(\mathfrak{U},\mathfrak{U}^{\dagger})=\int
d\phi\left\{i\frac{\mathfrak{U}^{\dagger}\partial_{\phi}\mathfrak{U}-\mathfrak{U}\partial_{\phi}\mathfrak{U}^{\dagger}}{2}+i\frac{\mathfrak{U}\mathfrak{U}-\mathfrak{U}^{\dagger}\mathfrak{U}^{\dagger}}{2}\frac{\partial_{\phi}\mathrm{E}_{\phi}}{2\mathrm{E}_{\phi}}-\mathfrak{H}\right\},
\end{equation}
where
\begin{equation}\label{4.11}
\mathfrak{H}=\left(\mathfrak{U}^{\dagger}\mathfrak{U}+\frac{1}{2}\right)\mathrm{E}_{\phi}
\end{equation}
is the Hamiltonian operator of the system. (\ref{4.11}) is also in analogy with a harmonic oscillator case, but in (\ref{4.10}) the additional term following from nonconstant character of ''the mass'' $E_\phi$ is present. A variational principle applied to ''the Universe action'' (\ref{4.10}) leads to the
Euler--Langrange equations of motion for the Universe annihilator $\mathfrak{U}$ and the Universe creator $\mathfrak{U}^{\dagger}$ as follows
\begin{equation}\label{4.11a}
i\partial_{\phi}\left[\begin{array}{c}\mathfrak{U}\\ \mathfrak{U}^{\dagger}\end{array}\right]=\left[\begin{array}{cc} \mathrm{E}_{\phi} & i\frac{\partial_{\phi}\mathrm{E}_{\phi}}{2\mathrm{E}_{\phi}}\\ i\frac{\partial_{\phi}\mathrm{E}_{\phi}}{2\mathrm{E}_{\phi}} & -\mathrm{E}_{\phi}\end{array}\right]\left[\begin{array}{c}\mathfrak{U}\\ \mathfrak{U}^{\dagger}\end{array}\right].
\end{equation}
The  evolutionary equations (\ref{4.11a}) can be understand as the manifestly inhomogeneous Heisenberg
equations of motion \cite{bog2} describing the one-dimensional global $\phi$-evolution of the
creator $\mathfrak{U}^{\dagger}$ and the annihilator $\mathfrak{U}$ on the appropriate Fock
space. The~~RHS of these equations is not equal to zero identically
as in the case of the usual Heisenberg equations of motion (See \emph{e.g.} Ref. \cite{pesk}). The nondiagonal terms results as an effect of
the evolutionary character the mass $\mathrm{E}_{\phi}$. In case of simple harmonic oscillator, when the mass is constant, the RHS vanishes
automatically and $\phi$-evolution is the Heisenberg type.

\subsection{\label{sec:6}Diagonalization Ansatz}

We see that both the integrand of (\ref{4.10}), that is in fact the Lagrangian of the secondary quantized quantum cosmology, and the equations of motion (\ref{4.11a}) are \emph{non-canonical} type. On the other words the $\phi$-evolution
of the Universe annihilator $\mathfrak{U}$ and the Universe creator $\mathfrak{U}^{\dagger}$ is not
Heisenberg type manifestly. Now we would like to obtain the Heisenberg
$\phi$-evolution; it is necessary condition for formulation of correctly defined quantum field theory associated to the quantum cosmology. In order to achieve one's goal we introduce \emph{ad hoc} the following ansatz\\

\noindent \textbf{Diagonalization Ansatz}
\textit{
Let us take into considerations the following assumptions
\begin{enumerate}
\item The operators $\mathfrak{U}^{\dagger}$ and $\mathfrak{U}$ on the appropriate Fock space for which the Lagrangian of the system is manifestly nondiagonal and
the global $\phi$-evolution is described by inhomogeneous Heisenberg
equations of motion, come into existence from the operators
$\mathtt{U}^{\dagger}$ and $\mathtt{U}$ due to the Bogoliubov
transformation
\begin{equation}\label{4.17}
\left[\begin{array}{cc}\mathfrak{U}^{\dagger}\\
\mathfrak{U}\end{array}\right]=\left[\begin{array}{cc}\mathrm{A}(\phi)&\mathrm{B}^{\ast}(\phi)\\
\mathrm{B}(\phi)&\mathrm{A}^{\ast}(\phi)\end{array}\right]\left[\begin{array}{cc}\mathtt{U}^{\dagger}\\
\mathtt{U}\end{array}\right],
\end{equation}
where the transformation matrix fulfills the rotation condition
\begin{equation}\label{4.18}
\det\left[\begin{array}{cc}\mathrm{A}(\phi)&\mathrm{B}^{\ast}(\phi)\\
\mathrm{B}(\phi)&\mathrm{A}^{\ast}(\phi)\end{array}\right]=1.
\end{equation}
\item The $\phi$-evolution of the operators $\mathtt{U}^{\dagger}$ and $\mathtt{U}$ is described by the Heisenberg equations of motion:
\begin{equation}\label{4.15}
i\partial_{\phi}\left[\begin{array}{c}\mathtt{U}\\ \mathtt{U}^{\dagger}\end{array}\right]=\left[\begin{array}{cc} \tilde{\mathrm{E}}_{\phi} & 0 \\ 0 & -\tilde{\mathrm{E}}_{\phi}\end{array}\right]\left[\begin{array}{c}\mathtt{U}\\ \mathtt{U}^{\dagger}\end{array}\right],
\end{equation}
with $\tilde{\mathrm{E}}_{\phi}$ as the diagonalization parameter. In general $\tilde{\mathrm{E}}_{\phi}$ is only immediate quantity and has non physical sense.
\item Canonically, in the operator basis $(\mathtt{U}^{\dagger},\mathtt{U})$ the Lagrangian of the system has diagonal form
\begin{equation}\label{4.14}
\mathrm{L}(\mathtt{U}^{\dagger},\mathtt{U})\equiv\mathrm{H}(\mathtt{U}^{\dagger},\mathtt{U})=\mathrm{E}_{\mathtt{U}}\mathtt{U}^{\dagger}\mathtt{U},
\end{equation}
where $\mathrm{E}_{\mathtt{U}}$ is the physical energy of the system.
\item The operator $\mathtt{N}=\mathtt{U}^{\dagger}\mathtt{U}$ has natural interpretation of the denisty of excitations number operator. $\mathtt{N}$ is an integral of motion
\begin{equation}\label{4.19}
\partial_{\phi}\mathtt{N}=0,
\end{equation}
and by this exist the \underline{stable} vacuum state $|0\rangle$ of
the system
\begin{equation}\label{4.20}
\mathtt{U}|0\rangle=0,~~\langle0|\mathtt{U}^{\dagger}=0
\end{equation}
\end{enumerate}}

From the Diagonalization Ansatz the following equations of motion for the coefficients
$\mathrm{A}$ and $\mathrm{B}$ in the Bogoliubov transformation arise
(\ref{4.17})
\begin{equation}\label{4.25}
\partial_{\phi}\left[\begin{array}{c}\mathrm{A}(\phi)\\ \mathrm{B}(\phi)\end{array}\right]=\left[\begin{array}{cc}i\mathrm{E}_{\phi}&\frac{\partial_{\phi}\mathrm{E}_{\phi}}{2\mathrm{E}_{\phi}}\\\frac{\partial_{\phi}\mathrm{E}_{\phi}}{2\mathrm{E}_{\phi}}&i\mathrm{E}_{\phi}\end{array}\right]\left[\begin{array}{c}\mathrm{A}(\phi)\\ \mathrm{B}(\phi)\end{array}\right],
\end{equation}
which with the initial values $\mathrm{A}(\phi(\tau_{0}))=1$ and $\mathrm{B}(\phi(\tau_{0}))=0$ can be integrated explicitly as
\begin{eqnarray}
\mathrm{A}(x)&=&\frac{1}{2x}e^{i\lambda(x^{3}-1)}\left[-x^{2}-1+2i\lambda\frac{x^{3}-1}{\ln{x}}(x^{2}-1)\right],\label{4.30}\\
\mathrm{B}(x)&=&\frac{1}{2x}e^{i\lambda(x^{3}-1)}(x^{2}-1),\label{4.31}
\end{eqnarray}
where $x=\frac{\phi(\tau)}{\phi(\tau_{0})}$ is dimensionless
parameter of evolution and
\begin{equation}\label{4.24a}
\lambda=\frac{16\pi
V}{9\mathcal{M}_{Pl}^{2}}\sqrt{\mathrm{H}_{Fields}(\tau)}\phi^{3}(\tau_{0}).
\end{equation}
The diagonalization parameter $\tilde{\mathrm{E}}_{\phi}$ has a form
\begin{equation}\label{4.23}
\tilde{\mathrm{E}}_{\phi}=-\frac{\lambda}{\phi(\tau_{0})}\left[3x^{4}-4x^{2}+\frac{x^{3}-1}{x^{3}\ln{x}}(x^{2}-1)^{2}+3\right]-i\frac{x^{4}-1}{2x^{3}},
\end{equation}
while the physical energy is
\begin{equation}\label{4.24}
\mathrm{E}_{\mathtt{U}}=-\frac{3\lambda}{2\phi(\tau_{0})}(x^{4}+1).
\end{equation}

\section{\label{sec:7}The Temperature}
Now we will give a thermodynamical interpretation of the above
results and, in particular, find a temperature associated with the
quantum system under consideration. From standard thermodynamics (Cf. \emph{e.g.} \cite{Jac,Huang}) one knows that the temperature of system $\mathrm{T}$ can be derived elementary, when one has computed entropy $\mathrm{S}$ and energy $\mathrm{E}$ of any physical system
\begin{equation}\label{T}
\mathrm{T}^{-1}=\left(\frac{\partial \mathrm{S}}{\partial
\mathrm{E}}\right)_{V}.
\end{equation}
According to the von Neumann approach, the entropy of system is a
vacuum expectation value as follows
\begin{equation}\label{S}
\mathrm{S}=k_{B}\langle\mathfrak{N}\ln{\mathfrak{N}}\rangle,
\end{equation}
where $\mathfrak{N}=\mathfrak{U}^{\dagger}\mathfrak{U}$ is the density of Universes number operator and $k_{B}$ is the Boltzmann constant. The energy of system can be established as
\begin{equation}\label{E}
\mathrm{E}=\langle\mathfrak{N}\mathfrak{H}\rangle.
\end{equation}
From (\ref{T}), (\ref{S}), (\ref{E}) it is clear that
\begin{eqnarray}\label{p}
\left(\frac{\partial \mathrm{S}}{\partial
\mathrm{E}}\right)_{V}&=&k_{B}\frac{\partial\langle\mathfrak{N}\ln{\mathfrak{N}}\rangle}{\partial\langle\mathfrak{N}\mathfrak{H}\rangle}
=k_{B}\frac{\partial_{\phi}\langle\mathfrak{N}\ln{\mathfrak{N}}\rangle}{\partial_{\phi}\langle\mathfrak{N}\mathfrak{H}\rangle}
\nonumber\\&=&k_{B}\frac{\langle\partial_{\phi}\mathfrak{N}\ln{\mathfrak{N}}+\partial_{\phi}\mathfrak{N}\rangle}{\langle\partial_{\phi}\mathfrak{N}\mathfrak{H}+\mathfrak{N}\partial_{\phi}\mathfrak{H}\rangle},
\end{eqnarray}
and now if one computes the vacuum expectation values in (\ref{p}) the formula can be received
\begin{equation}\label{p1}
\left(\frac{\partial \mathrm{S}}{\partial
\mathrm{E}}\right)_{V}=\frac{2k_{B}}{\mathrm{E}_{\phi}}\frac{1+\langle\ln{\mathfrak{N}}\rangle}{1+4\langle\mathfrak{N}\rangle+\left[2\langle\mathfrak{N}^{2}\rangle+\langle\mathfrak{N}\rangle\right]\frac{1}{\mathrm{E}_{\phi}}\frac{{\partial_{\phi}\mathrm{E}_{\phi}}}{\partial_{\phi}\langle\mathfrak{N}\rangle}}.
\end{equation}
Some elementary algebraic manipulations lead to the final result
\begin{equation}\label{T1}
\mathrm{T}=\frac{3\lambda}{8k_{B}\phi(\tau_{0})}\left[x^{5}-x^{3}+15x^{2}-6+\frac{3}{x^{3}}-\frac{15x^{2}}{x^{3}+4}-\frac{63}{4x^{3}(x^{3}+4)}\right].
\end{equation}
\section{\label{sec:8} Summary}
In this paper we have considered the homogenous cosmological model
and proposed quantization procedure for this model - the combination
of four well known rules of quantum mechanics and quantum field theory.
As a result we have obtained the quantum cosmology described in the
appropriate Fock space, that after using the proposed Diagonalization
Ansatz leads to correctly defined quantum field theory.
Of course, this is the only an example of this type quantum system, because of the homogenous cosmological metric is an
example of model of the generic Universe space-time. We stated the
quantum model under consideration is described in thermodynamical
terms and found the corresponding temperature. We saw that the
temperature of the system (\ref{T1}) is explicit function of the $x$ (or
redshift $z$).

In our opinion the further considerations in quantum cosmology can be studied in the proposed context. In the analogy to the results presented in this paper, it is very probable that proper development of the method will lead us to the rational model of quantum gravity, which will consistent with experimental data.

\section*{\label{sec:9}Acknowledgements}

The author benefited valuable discussions from Profs. I. Ya. Aref'eva, K. A. Bronnikov, and I. L. Buchbinder, and thanks to Prof. V. N. Pervushin for discussion in the context of his results presented in \cite{Perv}.


\begin{thebibliography}{99}
\bibitem{E} G. Esposito, \textit{Quantum Gravity, Quantum
Cosmology and Lorentzian Geometries,} Springer-Verlag, 1992.

\bibitem{W} S. Weinberg, \textit{Gravitation and Cosmology. Principles and Applications of the General Theory of Relativity}, Wiley, 1972.

\bibitem{Mis} Ch.W. Misner, K.S. Thorne, J.A. Wheeler \textit{Gravitation}, Freeman and Company, 1973.

\bibitem{Dir} P.A.M. Dirac \emph{Generalized Hamiltonian Dynamics}, {\it Proc. Roy.
Soc. (London)} {\bf A 246}, 326--332 (1958)

\bibitem{GT} D.M. Gitman, I.V. Tyutin, \textit{Quantization of Fields with Constraints}, Springer-Verlag, 1990.

\bibitem{WDW} J.A.~Wheeler, in \newblock{\em Batelle Rencontres: 1967,  Lectures in Mathematics and Physics,} edited by C. DeWitt and J.A. Wheeler , New York, 1968, p.~242

\bibitem{witt} B.C.~DeWitt, {\it Phys. Rev.} {\bf 160}, 1113 (1967).

\bibitem{bog1} N.N. Bogoliubov, D.V. Shirkov \textit{Introduction to the theory of quantized
fields} (in Russian), Nauka, 1984.

\bibitem{pesk} M.E. Peskin, D.V. Shroeder \textit{Introduction to quantum field theory}, Addison--Wesley, 1995.

\bibitem{bog2} N.N. Bogoliubov, A.A. Logunov, A.I. Oksak, I.T. Todorov \textit{General principles of quantum field theory} (in
Russian), FIZMATLIT, 2006.

\bibitem{B} I.L. Buchbinder \emph{Priv. comm.}

\bibitem{Jac} Z. Jacyna-Onyszkiewicz \textit{Principles of Quantum Thermodynamics} (in Polish), Wydawnictwo Naukowe UAM, Pozna\'n 1996.

\bibitem{Huang} K. Huang , \textit{Statistical Mechanics}, John Wiley \& Sons, Toronto 1987.

\bibitem{Perv} V.N. Pervushin, V.A. Zinchuk \emph{Phys. At. Nucl.} vol 70, No 3, 2007

\end{thebibliography}
\end{document}